\documentclass[a4paper,prb,superscriptaddress, amsfonts, amssymb, amsmath, reprint, showkeys, nofootinbib, twoside, twocolumn]{revtex4-2}
\usepackage[english]{babel}
\usepackage[utf8]{inputenc}
\usepackage{amsthm}
\usepackage{mathtools}
\usepackage{physics}
\usepackage{xcolor}
\usepackage{graphicx}
\usepackage[T1]{fontenc}
\usepackage[ pdftitle={Article}, pdfauthor={Author}]{hyperref} 
\usepackage{physics}
\usepackage[colorinlistoftodos, color=green!40, prependcaption]{todonotes}
\usepackage{dsfont}
\usepackage{chemformula}
\usepackage{units}
\usepackage{amsmath}
\usepackage{caption}
\usepackage{relsize}
\usepackage{enumitem}
\usepackage{epsfig}
\usepackage[amssymb]{SIunits}
\usepackage[normalem]{ulem}
\usepackage{esint}
\usepackage{multirow}

\begin{document}
\title{Non-local correlation effects due to virtual spin-flip processes in itinerant electron ferromagnets} 
\author{Sebastian Paischer} \email{sebastian.paischer@jku.at} 
\affiliation{Institute for  Theoretical Physics, Johannes Kepler  University Linz, Altenberger  Stra{\ss}e 69, 4040 Linz} 
\author{Giovanni Vignale}
\affiliation{The Institute for Functional Intelligent Materials (I-FIM), National University of Singapore, 4 Science Drive 2, Singapore 117544 }
\author{Mikhail I. Katsnelson} 
\affiliation{Institute for Molecules and Materials, Radboud University, Heyendaalseweg 135, 6525AJ Nijmegen, The Netherlands} 
\author{Arthur Ernst}
\affiliation{Institute for Theoretical Physics, Johannes Kepler University Linz, Altenberger  Stra{\ss}e 69, 4040 Linz}
\affiliation{Max Planck Institute of Microstructure Physics, Weinberg 2, D-06120 Halle, Germany}
\author{Pawe\l{} A. Buczek}
\affiliation{Department of Engineering and Computer Sciences, Hamburg  University of Applied Sciences, Berliner Tor 7, 20099 Hamburg, Germany} 

\date{\today}

\begin{abstract}
We present an {\em ab initio} method for electronic structure calculations, which accounts for the interaction of electrons and magnons in ferromagnets. While it is based on a many body perturbation theory we approximate numerically complex quantities with quantities from time dependent density functional theory. This results in a simple and affordable algorithm which allows us to consider more complex materials than those usually studied in this context ($3d$ ferromagnets) while still being able to account for the non-locality of the self-energy. Furthermore, our approach allows for a relatively simple way to incorporate self-consistency. Our results are in a good agreement with experimental and theoretical findings for iron and nickel. Especially the experimental exchange splitting of nickel is predicted accurately within our theory. Additionally, we study the halfmetallic ferromagnet NiMnSb concerning its non-quasiparticle states appearing in the bandgap due to spin-flip excitations.
\end{abstract}


\maketitle

\section{Introduction}

Even after the discovery of quantum mechanical exchange interactions as the reason for magnetic
ordering, itinerant-electron magnetism including ferromagnetism of elemental Fe, Co,
and Ni remained a challenging problem for decades \cite{Vonsovsky1974,Moriya2012,Kubler2017}.
The main issue is a coexistence of band (itinerant) and atomic (localized) features in
the behavior of $3d$-electrons in solids, something difficult to combine in a single
theory \cite{Vonsovsky1989}. The density functional theory (DFT), which was for a long time a method by default for a quantitative description of electronic properties of solids was applied to 
itinerant-electron magnets with great success \cite{Kubler2017} but atomic-like
features such as remnants of atomic multiplet structure in some transition metal compounds
were not easy to take into account. The idea to combine the DFT with the dynamical mean-field
theory (DMFT) \cite{Anisimov1997,Lichtenstein1998,Katsnelson1999,Kotliar-DMFT} was an
essential step to solve the problem. Ferromagnetic transition metals were one of the
aims of this approach from the very beginning \cite{Katsnelson1999,Katsnelson2000,
Lichtenstein2001} and the results were very promising, especially in the description
of their high-temperature magnetism and in the successful description of the famous 6 eV
satellite in Ni \cite{Lichtenstein2001}. Furthermore detailed calculations and
comparison with the experimental data on angle-resolved photoemission
\cite{Braun2006,Grechnev2007,SanchezBarriga2009,SanchezBarriga2012} demonstrated
that whereas the combination of the DFT and the DMFT seems to be sufficient for a
quantitative description of Ni there are still remaining problems with Fe,
despite an essential improvement of the description in the DFT+DMFT in comparison with the
pure DFT approach. In particular, the DMFT being a local approximation predicts a
wavevector-independent renormalization of effective masses with respect to the
bare DFT values. Experimentally, it works quite accurate for Ni, a bit worse for
Co and not accurate for Fe \cite{SanchezBarriga2012}. 

Another important group of itinerant-electron ferromagnets are weak itinerant
ferromagnets such as ZrZn$_2$ \cite{Moriya2012} and half-metallic ferromagnets
such as several Heusler alloys or CrO$_2$ \cite{Katsnelson2008}. The latter group
of materials is especially interesting since for them many-body effects result
not only in a renormalization and damping of electron quasiparticle spectrum but
in appearance of a qualitative new feature, non-quasiparticle (or spin-polaron) states 
(NQPS) in a majority or minority electron energy gap \cite{Edwards1973,Irkhin1990,Katsnelson2008}.
These states were experimentally found in full Heusler alloy Co$_2$MnSi \cite{Chioncel2008}
and in CrO$_2$ \cite{Fujiwara2017}. The DFT+DMFT calculations can successfully
describe NQPS at least at a qualitative level \cite{Chioncel2003,Katsnelson2008}. 
At the same time, an appropriate quantitative description is possible only beyond the local approximation for the electron self-energy, and in particular, 
it requires a correct description of the magnon dispersion \cite{Edwards1973,Irkhin1990}.\\
An important type of many-body effects in itinerant-electron magnets
originates from the interaction of electrons with bosonic spin-flip excitations,
both coherent (magnons) and incoherent (Stoner particle-hole) excitations \cite{Moriya2012}. While there has been a steady progress in understanding the
 properties of spin-flip excitations at a model level only little is known about
 microscopic details of their interactions with the electronic
 degrees of freedom in specific materials. The conventional DFT as well as the GW \cite{Hedin1965,GW_review}
 do not take these processes into account. The DFT+DMFT does take them into account
 but only in a local approximation whereas wavevector-dependence of the corresponding
 contributions to the electron self-energy can be quite important, as we discussed
 above for the case of Fe. At the same time, the coupling has been directly experimentally detected
 \cite{Hofmann2009,Prokop2009,Zakeri2014,Balashov2006} and is
 believed to crucially influence properties of magnetic
 materials leading to a number of remarkable phenomena.  The latter
 include spin-dependent lifetime of excited electronic states
 \cite{Schmidt2010} and inelastic electron transport
 \cite{Balashov2008}. Consequently, also the mean free path of
 excited electrons is spin dependent 
 \cite{Hong1999,Zhukov2004,Zhukov2004a}.  The electron-magnon interaction is also 
 discussed as the moving force for the formation of
 Cooper pairs in certain high-temperature superconductors, in
 particular in the pnictide and cuprate families \cite{Mazin2008,Essenberger2012,Essenberger2014,Essenberger2016}. Furthermore, ultrafast magnetization
 switching phenomena in solids are believed to involve the coupling
 between the electronic and spin degrees of freedom as well
 \cite{Wilson2017,ElGhazaly2019}.  Last but not least, the coupling
 of magnons and electrons contributes to the renormalization of the
 electronic bands and influences the exchange splitting
 \cite{Sasioglu2010}. For the particular case of half-metallic
 ferromagnets, the effects of the electron-magnon interaction
 are systematically reviewed in Ref.\cite{Katsnelson2008}.

 Apart from the DFT+DMFT approach discussed above, the electron-magnon
 interaction can be treated within the many body perturbation theory (MBPT)
 \cite{Zhukov2005,Friedrich2019}. The MBPT relies on a
 perturbation expansion of the electronic self-energy based on the 
 general Hedin equations \cite{Hedin1965}. MBPT calculations are computationally quite expensive, hence until recently only results from
 simple bulk systems are available \cite{Mueller2019,Nabok2021}.

 The approach presented in this work is based on the MBPT but avoids
 the numerical complexity by approximating some complex quantities
 with quantities from the time dependent DFT (TDDFT) \cite{Buczek2011,Buczek2011a,Sandratskii2012}, which are much
 easier to calculate. This allows to compute the influence of the
 electron-magnon interaction on the electronic structure for complex
 materials like half-metallic ferromagnets from first principles. This
will allow us to consider NQPS taking into account non-locality
of the spin-flip contributions to the electron self-energy and
thus going beyond the DFT+DMFT treatment of these states \cite{Chioncel2003,
Katsnelson2008}. Most recently, a method similar to ours including electron-magnon interactions based on MBPT was put forth \cite{Nabok2021}. However, while we opted to minimize the numerical burden of calculations (by using quantities from TDDFT and hence being able to account for more complex systems), they also included the GW term in their calculations leading to clearly improved results for the elemental $3d$ ferromagnets.
We put forth another physically motivated simplification leading to a reduced computational cost of our calculations. In many cases, upon including the interaction, practically the entire spectral density of spin-flip processes is shifted from the Stoner energy window into the energy range of spin-waves \cite{Buczek2011}. This suggests that, unless the Landau damping itself is the subject of the investigation, it might be possible to replace the dynamical susceptibility with its ``Heisenberg counterpart''. We refer to this approach as ``magnon-pole approximation''. Its validity is discussed in detail and proven later in this paper.

 So far, our method is a one-shot method, meaning that we calculate
 the electronic structure of the materials in question and
 consequently perform the MBPT based algorithm to compute the 
 self-energy representing the influence of the electron-magnon
 scattering. Van Schilfgaarde and coauthors \cite{Schilfgaarde2006} showed 
 with a similar method, combining the DFT and the GW theory, that the results of
 this approach can depend on the starting electronic structure. 
It is shown in this work that one should ideally aim to use
 the results of the MBPT to construct a self-consistent cycle, making
 the final result independent of the starting Green function. Contrary to other methods based on the MBPT \cite{Mueller2019,Nabok2021}, our approach offers a relatively simple way to incorporate a self-consistency. After the calculation of the electron-magnon self-energy, the renormalized Green function can be used to calculate new Heisenberg exchange parameters utilized in the magnon-pole approximation. Consequently, a new self-energy can be calculated. The implementation of this numerical scheme is a current work in progress.
 
 

\begin{figure*}
    \includegraphics[width=\textwidth]{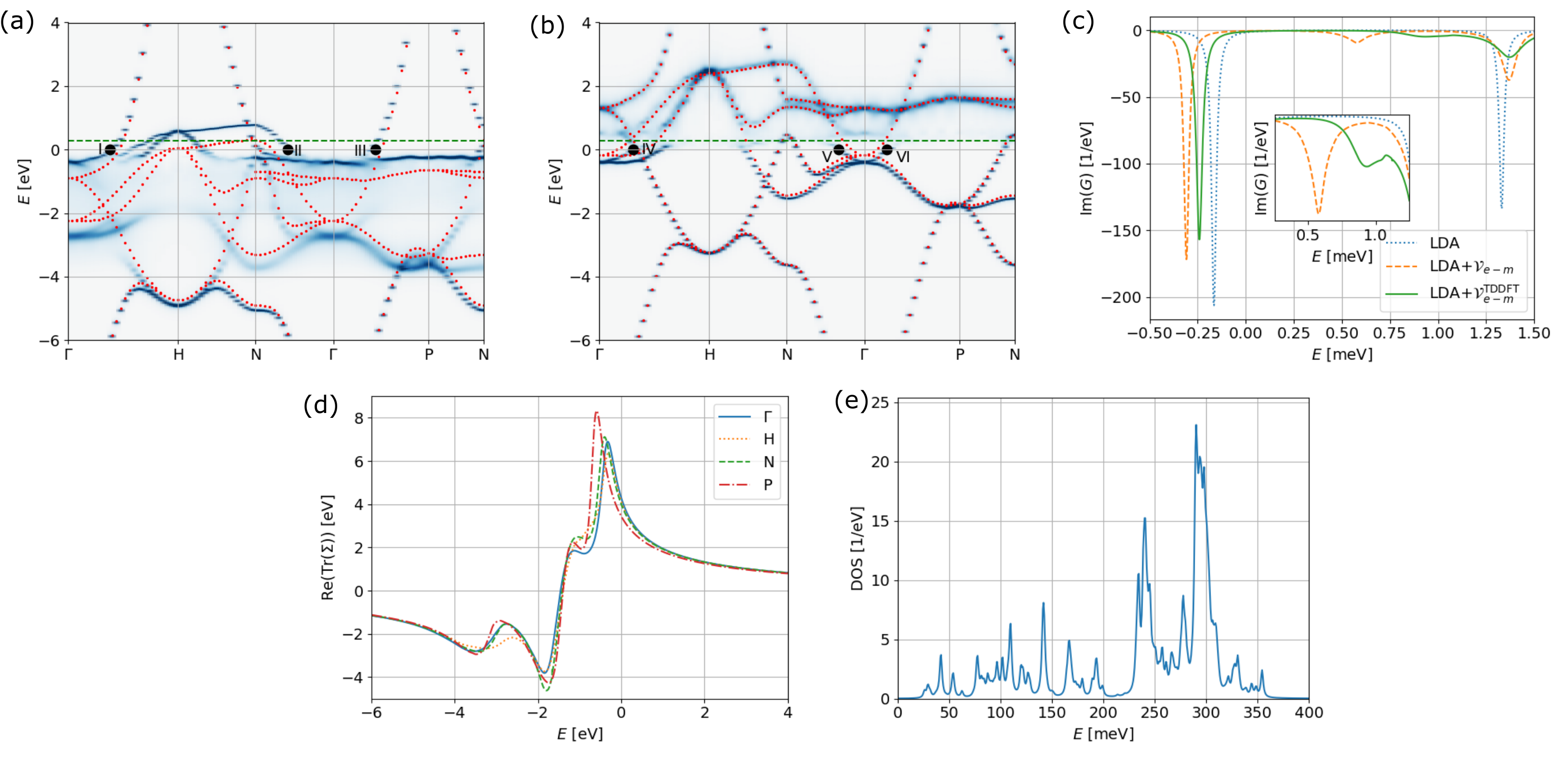}
    \caption{Effect of the electron-magnon interaction in bcc iron. (a) Electronic band structure of majority carriers under the influence of the electron-magnon interaction. The red dots represent the energy levels within the LDA while the blue background indicates the renormalized spectrum. The slope of the bands marked with black dots were compared to ARPES measurements in table \ref{tab_mass2}. The renormalized Fermi energy is represented by the dashed green line. (b) Same as (a) for minority carriers. (c) Comparison of the spectral functions obtained with (LDA+$\mathcal{V}_{e-m}$) and without (LDA+$\mathcal{V}_{e-m}^{\text{TDDFT}}$) the magnon-pole approximation at the $\Gamma$ point for minority carriers. (d) Real part of the trace of the self-energy for different points in the Brillouin zone. (e) Magnonic density of states.}
    \label{fig_Fe}
\end{figure*}

\section{Theory}\label{chap_theory}
\subsection{Many body perturbation theory}
Magnons are magnetic excitations, i.e. transitions to an excited state
with a different value of the total spin projection. We will use this
term for both coherent collective excitations (spin waves) and for
incoherent spin-flip electron-hole pairs known as Stoner excitations
\cite{Vonsovsky1974,Moriya2012,Kubler2017}. Spin-waves are transverse 
fluctuations of magnetization that may be intuitively understood as a 
correlated precession of atomic magnetic moments. In the
many body language they are formed due to the multiple scattering of
particle and hole of opposite spin projection which are described,
in a minimal set, as a summation of ladder diagrams
\cite{Vignale2005,Sasioglu2010}.

The interaction between magnons and electrons can be formulated in
terms of an electron self-energy. Within the Hedin equations \cite{Hedin1965}
the general expression for the self-energy reads 
\begin{align}\label{eqn_slfe_Hedin}
    \includegraphics[width=0.25\textwidth]{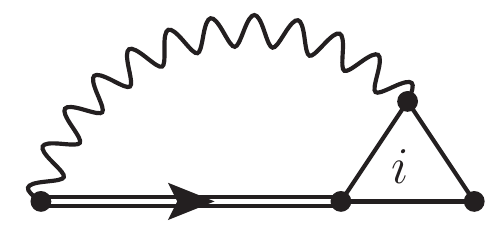}
\end{align}
where the screened interaction $W$ is drawn as wavy line, the many-body
Green function $G$ is represented by a double line and the three-leg vertex $\varGamma$ (drawn as triangle)is defined through the self-consistent equation 
\begin{align}\label{eqn_vertex}
    \includegraphics[width=0.45\textwidth]{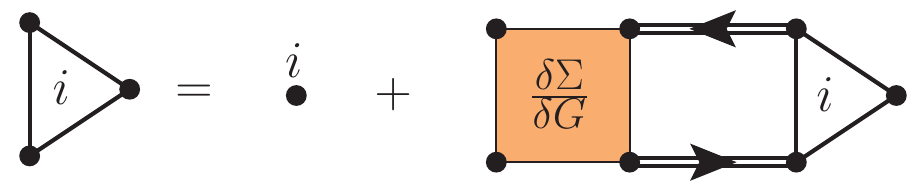}
\end{align}
with the effective proper two-particle interaction (four-leg vertex) block given by the functional derivative
\begin{align}
    \fdv{\varSigma_{\alpha\beta}(\mathit{1},\mathit{2})}{G_{\mu\nu}(\mathit{4},\mathit{3})}.
\end{align}  
In the latter expression the italic numbers represent the spatial and
time variable $\mathit{1}=(\vb*{x},t)$. \newline
From now on we consider only ferromagnetic systems where charge fluctuations are decoupled from the
transverse spin fluctuations. At the same time, they are coupled with
longitudinal spin fluctuations \cite{Katsnelson2004,Buczek2020} but we will not
consider these processes in the present paper. As long as the fundamental interaction (Coulomb potential in our case)
is spin independent, even in the case of spin polarized systems, the
screened interaction $W$ cannot mediate a spin flip. Moreover, the
screened interaction depends only on the charge part of the vertex
function and therefore it is not affected by the electron-magnon
interaction at all.  Consequently, the only way for the
electron-magnon scattering to enter the theory is through the vertex
correction in the self-energy relation \ref{eqn_slfe_Hedin}. We use in
our theory the three-leg vertex to first order
\begin{align}
    \includegraphics[width=0.4\textwidth]{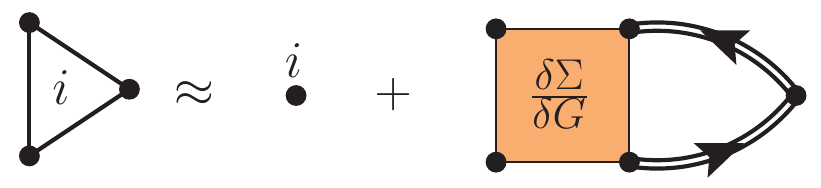}
\end{align}
The first term in the
latter equation, leading to the well known GW approximation
\cite{Hedin1965,GW_review}, does not involve electron-magnon scattering. 
Hence, for the remainder of this work, only the second term (linear in $\fdv{\varSigma}{G}$) 
will be analyzed. As was shown in a number of preceding works
\cite{Ng1986,Vignale1985b}, $\fdv{\varSigma}{G}$ can be divided into three classes
of which only one type (diagrams reducible in the electron-hole
channel) involves the scattering of an electron with a hole of opposite
spin (i.e. involve magnons). Those diagrams are of the form
\begin{align}\label{eqn_Xi}
	\includegraphics[width=0.3\textwidth]{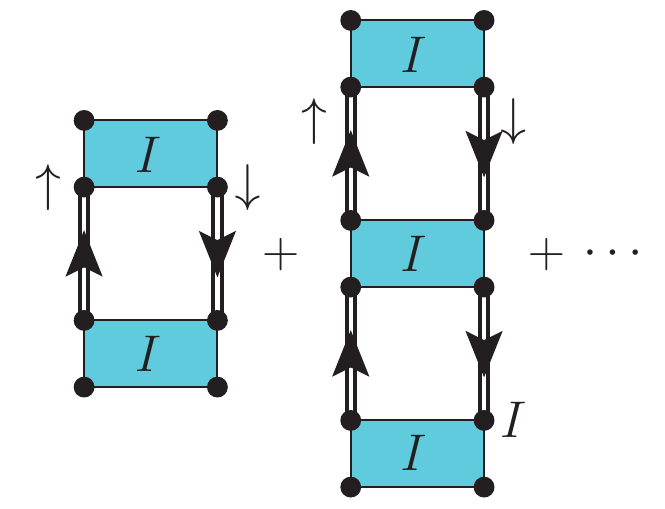}
\end{align}
where $I$ are irreducible diagrams and the arrows represent the spin
directions. 
The latter diagrams can be written compactly using
the transverse two particle correlation function $L$ defined through
the Bethe-Salpeter equation
\begin{align}
	\includegraphics[width=0.45\textwidth]{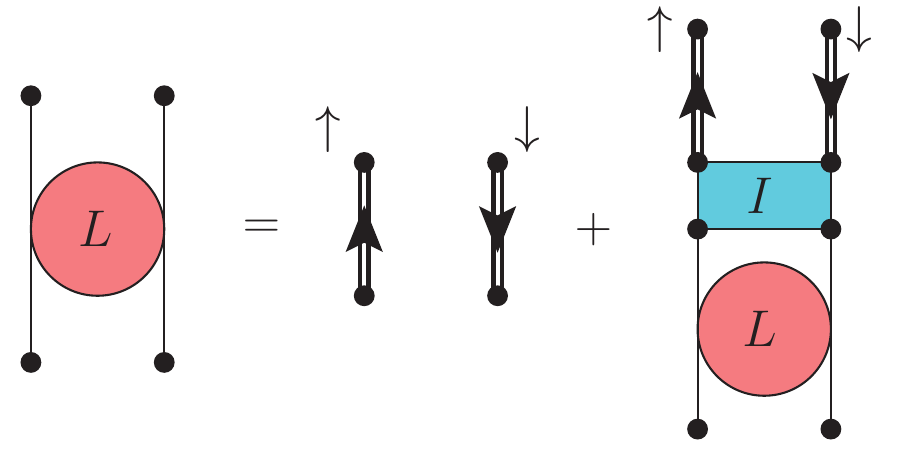}.
\end{align}
This allows to write the  electron-magnon self-energy (often called GT contribution) as

\begin{align}\label{eqn_slfe}
    \includegraphics[width=0.22\textwidth]{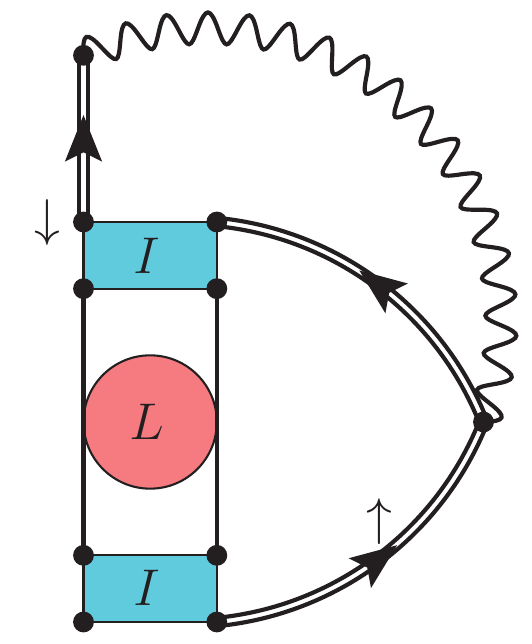}.
\end{align}
The main drawback of the latter equation is its high numerical cost,
especially for complex systems. Hence, more approximations are in order. First, $I$ is assumed to
be local
\begin{align}
	I_{\alpha\beta\mu\nu}(\mathit{1},\mathit{2},\mathit{3},\mathit{4})&\approx I_{\alpha\alpha\mu\mu}(\mathit{1},\mathit{1},\mathit{3},\mathit{3})\delta(\mathit{1}-\mathit{2})\delta(\mathit{3}-\mathit{4}).
\end{align}
Similar approximations have been used successfully to calculate spin
wave spectra \cite{Zhukov2004,Aryasetiawan1999,Sasioglu2010}.  Second, we also
approximate the extra $W$ in equation \ref{eqn_slfe} with
$I$. This leads to a geometric series which can be
summed. Furthermore, we make one more approximation in the spirit of
Ng \& Singwi \cite{Ng1986} in starting the summation one term earlier. This
allows to write the self-energy in terms of the susceptibility 
\begin{align}
	&\varSigma_{\alpha\beta}(\mathit{1},\mathit{2})\approx\nonumber\\	&\text{i}\sigma^i_{\alpha\eta}I^{ij}(\mathit{1},\mathit{1},\mathit{4},\mathit{4})\chi^{jm}(\mathit{4},\mathit{6})I^{mn}(\mathit{6},\mathit{6},\mathit{2},\mathit{2})\sigma^n_{\eta'\beta}G_{\eta\eta'}(\mathit{1},\mathit{2})
\end{align}
as the susceptibility is given by the contraction of the two particle correlation function
\begin{align}\label{eqn_chi_L}
	\chi^{ij}(\mathit{1},\mathit{2})=-\text{i}\sigma^i_{\alpha\beta}\sigma^i_{\gamma\delta}L_{\alpha\beta\gamma\delta}(\mathit{1},\mathit{1}^+,\mathit{2}^+,\mathit{2}).
\end{align}
In the latter equations we used a summation and integration convention: Each spatial or time variable (spin index) appearing only on one side of an equation is integrated (summed) over. The plus as superscript
indicates that the time variable is modified by an infinitesimal $\epsilon$:
$\mathit{1}^+=(\vb*{x},t+\epsilon)$.

\subsection{Time dependent density functional theory}
In order to arrive at a feasible
computational scheme, we approximate the complex quantities from the many
body perturbation theory by quantities from the time dependent
density functional theory. The proper susceptibility can be obtained from the linear response
TDDFT by means of the susceptibility Dyson equation \cite{Katsnelson2004,Buczek2011}
\begin{align}\label{eqn_susc_dyson}
	&\chi^{ij}_{\text{TDDFT}}(\vb*{x}_1,\vb*{x}_2,E)=\chi_{\text{KS}}^{ij}(\vb*{x}_1,\vb*{x}_2,E)+\nonumber\\
	&\sum_{m,n}\iint\dd[3]{x_3}\dd[3]{x_4}\chi_{\text{KS}}^{im}(\vb*{x}_1,\vb*{x}_3,E)\nonumber\\
	&\left(K_{\text{xc}}^{mn}(\vb*{x}_1,\vb*{x}_2,E)+\frac{2\delta_{m0}\delta_{n0}}{\abs{\vb*{x}_3-\vb*{x}_4}}\right)\chi^{nj}_{\text{TDDFT}}(\vb*{x}_4,\vb*{x}_2,E)
\end{align}
where the Kohn-Sham (KS) susceptibility is given by
\begin{align}\label{eqn_KS_susc}
	\chi_{\text{KS}}^{ij}(\vb*{x}_1,\vb*{x}_2,E)=\sum_{mn}\sigma^i_{\alpha\beta}\sigma^j_{\gamma\delta}(f_m-f_n)\nonumber\\
	\frac{\phi_{m\alpha}(\vb*{x}_1)^\star \phi_{n\beta}(\vb*{x}_1)\phi_{n\gamma}(\vb*{x}_2)^\star \phi_{m\delta}(\vb*{x}_2)}{E-(E_{m\alpha}-E_{n\beta})+i\epsilon}
\end{align}
with the Fermi-Dirac function $f_n$ of the KS state $\phi_{n\alpha}$
at the energy $E_{n\alpha}$.  In practice, the KS susceptibility and the exchange correlation kernel $K_{\text{xc}}$ are computed interdependently from each other and, when combined in the susceptibility Dyson equation, yield slight deviation from the perfect Goldstone mode due to numerical inaccuracies. In our case, this deviation is of the order of few meV and can be removed (corrected) completely by imposing a suitable sum-rule on $K_{\text{xc}}$. This issues were discussed carefully in \cite{Buczek2011} and \cite{Mueller2016}. \\
To be fully consistent at this point, one
also needs to replace $I$ with the exchange correlation
kernel $K_{\text{xc}}$ and replace the many body Green function with
the KS Green function used for the calculation of the KS
susceptibility (equation \ref{eqn_KS_susc}). This can be seen by
comparison of equation \ref{eqn_susc_dyson} with the expression for
the susceptibility derived in the framework of the MBPT \ref{eqn_chi_L}
when $I$ is assumed to be local. 
In this work we only use the exchange correlation kernel from the
adiabatic local spin density approximation (ALSDA) \cite{Katsnelson2004}. 
This leads to the expressions
\begin{align}\label{eqn_slfe_final}
\varSigma_\uparrow(\mathit{1},\mathit{2})=&2\text{i}K_\text{xc}^{+-}(\mathit{1})\chi^{-+}_{\text{TDDFT}}(\mathit{1},\mathit{2})K_\text{xc}^{+-}(\mathit{2})G_{\downarrow}(\mathit{1},\mathit{2})\nonumber\\				
\varSigma_\downarrow(\mathit{1},\mathit{2})=&2\text{i}K_\text{xc}^{-+}(\mathit{1})\chi^{+-}_{\text{TDDFT}}(\mathit{1},\mathit{2})K_\text{xc}^{-+}(\mathit{2})G_{\uparrow}(\mathit{1},\mathit{2}),
\end{align}
which can also be written as
\begin{align}\label{eqn_GVeff_TDDFT}
	\varSigma_\uparrow(\mathit{1},\mathit{2})=&\text{i}\mathcal{V}^{-,\text{TDDFT}}_{e-m}(\mathit{1}^+,\mathit{2})G_\downarrow(\mathit{1},\mathit{2})\nonumber\\
	\varSigma_\downarrow(\mathit{1},\mathit{2})=&\text{i}\mathcal{V}_{e-m}^{+,\text{TDDFT}}(\mathit{1}^+,\mathit{2})G_\uparrow(\mathit{1},\mathit{2})
\end{align}
where
\begin{align}
	\mathcal{V}^{-,\text{TDDFT}}_{e-m}(\mathit{1},\mathit{2})&=2K_\text{xc}^{+-}(\mathit{1})\chi_{\text{TDDFT}}^{-+}(\mathit{1},\mathit{2})K_\text{xc}^{+-}(\mathit{2})\nonumber\\
	\mathcal{V}^{+\text{TDDFT}}_{e-m}(\mathit{1},\mathit{2})&=2K_\text{xc}^{-+}(\mathit{1})\chi_{\text{TDDFT}}^{+-}(\mathit{1},\mathit{2})K_\text{xc}^{-+}(\mathit{2})
\end{align}
represents a magnon mediated effective interaction. Analogous expressions have been derived in several reports using different theoretical approaches \cite{Vignale1985b,Ng1986,Schweflinghaus2014,Bouaziz2020}. Finally, the self
energy is used to renormalize the electronic Green function obtained
from the local density approximation (LDA) by means of the Dyson
equation 
\begin{align}
	G_{\alpha\beta}(\mathit{1},\mathit{2})=G^0_{\alpha\beta}(\mathit{1},\mathit{2})+G^0_{\alpha\mu}(\mathit{1},\mathit{3})\varSigma_{\mu\nu}(\mathit{3},\mathit{4})G_{\nu\beta}(\mathit{4},\mathit{2})
\end{align}
with the LDA Green function $G^0$.
The use of quantities readily
available in the TDDFT substantially reduces the numerical burden,
comparing to analogous expressions involving the screened Coulomb
interaction
\cite{Zhukov2004,Sasioglu2010,Aryasetiawan1999,Springer1998,Karlsson2000}
and allows to address large systems like NiMnSb and in the future two
dimensional heterostructures.\newline
In practice, the calculation of the electron-magnon interaction with the TDDFT susceptibility is still problematic as the susceptibility is needed for an immense number of energy points. To reduce the number of points the fluctuation-dissipation theorem  can be utilized \cite{Vignale2005}. After the calculation the spectral weights 
\begin{align}
	\mathcal{C}^\pm_{\text{TDDFT}}(\vb*{r},\vb*{r}',\vb*{k},E)=&\text{i}\theta(E)\left[\chi^\pm_{\text{TDDFT}}(\vb*{r},\vb*{r}',\vb*{k},E+\text{i}\epsilon)-\right.\nonumber\\
    &\left.\chi^\pm_{\text{TDDFT}}(\vb*{r}',\vb*{r},\vb*{k},E+\text{i}\epsilon)^\star\right]
\end{align}
along a path parallel to the real axis, the TDDFT susceptibility
at arbitrary energies in the complex plane can be calculated by
simple integration of the spectral representation of the susceptibility
\begin{align}\label{eqn_chi_spectral}
	\chi^\pm_{\text{TDDFT}}(\vb*{r},\vb*{r}',\vb*{k},E)=&\int\frac{\dd{E'}}{2\pi}\frac{\mathcal{C}^\pm(\vb*{r},\vb*{r}',\vb*{k},E')}{E-E'+\text{i}\epsilon}\nonumber\\
    &-\int\frac{\dd{E'}}{2\pi}\frac{\mathcal{C}^\mp(\vb*{r}',\vb*{r},\vb*{k},E')}{E+E'-\text{i}\epsilon}
\end{align}
However, due to the high numerical cost of the TDDFT susceptibility, this approach is only viable for fixed $\vb*{k}$ and energies close to the Fermi energy. Hence, the magnon-pole approximation (discussed in the chapter \ref{chap_magnon-pole}) was used for all following results unless otherwise noted.\newline\newline
With the magnon propagator $\mathcal{V}_{e-m}$ playing the role of the
effective interaction, equations \ref{eqn_GVeff_TDDFT} convey a
compelling physical picture of the electron-magnon scattering in
solids. Electrons (holes) in a given state can emit a magnon and decay
into another electron (hole) state of opposite spin providing 
energy and momentum differences between the initial and the final particle
states are accounted for by this magnon.
Furthermore, due to the prevalence of up-to-down magnon processes over
the down-to-up ones, in strong ferromagnets, effectively, only down
electrons and up holes interact with the spin-flip excitations,
respectively loosing and gaining energy in this process.
The above observations allow us to justify the deployment of the TDDFT
quantities in the MBPT scheme. 

A magnon corresponds to the transverse
fluctuations of the magnetization $\delta m$. In the spirit of the
TDDFT, these fluctuations give rise to the fluctuating
exchange-correlation potential (the transverse magnetic field in the
case of a collinear ground state and the ALSDA) while the potential is given by the
exchange-correlation kernel $K_{\text{xc}}$, such that $\delta
B^\pm_{\text{xc}} = K_{\text{xc}} \delta m^\pm$. We note that the very
same relationship governs calculations of the enhanced
susceptibility in the susceptibility Dyson equation,
eq. \ref{eqn_susc_dyson}. Thus, the abstract concept of
electron-magnon scattering is reinterpreted as the interaction of the
electron with the space- and time-dependent fluctuating exchange-correlation field associated with the
magnon \cite{Kukkonen1979}. In the case of the TDDFT this field is
straightforwardly given as $K_{\text{xc}} \delta m^\pm$ where
$K_{\text{xc}}$ naturally plays the role of the interaction vertex in
our equations. Recall that equivalent interpretation appears in
quantum electrodynamics \cite{Feynman1998} where the interaction of an
electron and photon can be understood as the interaction of the
electron with the electromagnetic field corresponding to this photon.

Let us, however, recall that the equations \ref{eqn_GVeff} are
obtained upon including only the minimal set of diagrams involving
scattering of electron-hole pairs of opposite spins. Therefore, while
the exchange of virtual magnons appears to be a dominating
contribution of spin fluctuations on the band structure in itinerant
magnets, it is necessarily only an approximate, albeit compelling
model.

\subsection{Magnon-pole approximation}\label{chap_magnon-pole}
While the Kohn-Sham susceptibility (equation \ref{eqn_KS_susc})
consists of a broad spectrum of spin flip excitations located at the
energy scale of the exchange splitting, the TDDFT
susceptibility (equation \ref{eqn_susc_dyson}) develops new low energy
singularities corresponding to spin waves, i.e. coherent precession of
the atomic magnetic moments. Interestingly, in most cases, practically
the entire spectral density is shifted from the Stoner energy window
into the energy range of the spin-waves. This suggests that, unless
the Landau damping itself is the subject of the investigation, it
might be possible to replace the dynamical susceptibility by its
''Heisenberg counterpart`` $\chi^H$ with singularities located at the
undamped spin-wave energies and spatial dependence corresponding to
the rigid tilt of the atomic magnetic moments $\mu$:
\begin{align}
	\left[\chi_{\text{H}}^{-1}\right]_{ij}=(2g\mu_i)^{-1}\left(z\delta_{ij}+g\mu_j^{-1}J_{ij}-g\delta_{ij}\mu_j^{-1}\sum_\ell J_{i\ell}\right)
	\label{exchh}
\end{align}
Here, the exchange interactions $J$ were calculated from the first
principles magnetic force theorem \cite{Lichtenstein1987} and the
Lande-factor $g\approx2$ was used. The relation between the full expression
for the dynamical susceptibility and its Heisenberg-like form (\ref{exchh})
is discussed in detail in Ref. \cite{Katsnelson2004}. We show that in many cases the
use of this Heisenberg-like susceptibility instead of the TDDFT
susceptibility in the expression for the self-energy changes the
results only marginally. At the same time, as computations of the TDDFT
susceptibility are computationally demanding, this
''magnon-pole-approximation`` results in a relatively inexpensive
numerical scheme. We also note that the magnon-pole approximation is
equivalent to the assumption of a strong ferromagnet. Using this approximation, the self-energy is given by
\begin{align}\label{eqn_GVeff}
	\varSigma_\uparrow(\mathit{1},\mathit{2})=&\text{i}\mathcal{V}^{-}_{e-m}(\mathit{1}^+,\mathit{2})G_\downarrow(\mathit{1},\mathit{2})\nonumber\\
	\varSigma_\downarrow(\mathit{1},\mathit{2})=&\text{i}\mathcal{V}_{e-m}^{+}(\mathit{1}^+,\mathit{2})G_\uparrow(\mathit{1},\mathit{2})
\end{align}
where
\begin{align}
	\mathcal{V}^{-}_{e-m}(\mathit{1},\mathit{2})&=2K_\text{xc}^{+-}(\mathit{1})\chi_{\text{H}}^{-+}(\mathit{1},\mathit{2})K_\text{xc}^{+-}(\mathit{2})\nonumber\\
	\mathcal{V}^{+}_{e-m}(\mathit{1},\mathit{2})&=2K_\text{xc}^{-+}(\mathit{1})\chi_{\text{H}}^{+-}(\mathit{1},\mathit{2})K_\text{xc}^{-+}(\mathit{2}).
\end{align}


\subsection{Goldstone mode, double counting and self-consistency}
We conclude this section with a short comment on double counting which
is a typical problem in approaches combining DFT with more
sophisticated methods like the dynamic mean field theory or the MBPT
\cite{Grechnev2007,Mueller2019}. The problem arises on two different
levels in methods presented in the literature. First, the introduction
of two different MBPT terms can lead to double counting as discussed
in \cite{Mueller2019}. The authors include (part of) the GW term
\cite{Mueller2016} for their susceptibility to restore the correct
Goldstone mode and then study the impact of the electron-magnon
interaction with the same diagrams we use in this work. This
leads to double counting if the diagram of second order in
$I$ is included in the calculation of the electron-magnon
scattering self-energy (see the text above equation
\ref{eqn_slfe_final}). In fact, the second order term alone is unphysical as it already includes double counting \cite{Mueller2019}. To strictly avoid double counting, one has to start from the third order in $I$ by subtracting the second order term from the geometrical series of ladder diagrams. However, it was shown in \cite{Mueller2019,Nabok2021} that this leads to the violation of the causality manifesting itself in the wrong sign of the imaginary part of the self-energy. This, in turn, can be corrected by the inclusion of numerically expensive GW term itself \cite{Nabok2021}, which, however, does not contribute to the spin-dependent electron scattering and is not considered in this work. For the materials studied in this work, the electron-magnon interaction exerts essentially the same effect on the electronic structure as in \cite{Mueller2019} and DMFT methods \cite{Katsnelson1999,Braun2006,Grechnev2007,SanchezBarriga2009,SanchezBarriga2012}. Hence, the double counting issue does not seem to be decisive here. Furthermore, in our case the exclusion of the second order term would also lead to a significant increase in the numerical cost of our scheme (especially in combination with the
magnon-pole approximation discussed below). Hence, we choose to include the second order diagram. As discussed in section II.B, the resulting full geometric series of ladder diagrams corresponds to a compelling physical picture of an electron interacting with the fluctuating Kohn-Sham potential associated with the magnon. Concerning the fact that your scheme is not self-consistent and we take the Kohn-Sham band structure as the starting point, we find the approximation physically acceptable.\\
Second, double counting may arise due to the fact
that it is unknown which many body effects are already included in the
exchange correlation potential of the DFT. While the double counting of the
first kind can be taken into account exactly (at least in principle), this
is not the case for the second kind. In the literature these double
counting issues are often assumed to be small due to the fact that the DFT
potentials are based upon the homogeneous electron gas which includes
only little correlation effects at the relevant densities
\cite{Mueller2019}. Other publications utilize a heuristic double
counting correction \cite{Grechnev2007}. In this work we resort to the
first option and assume the double counting corrections are small
enough to neglect them. \newline\newline

 Finally, we note that our method is a one-shot scheme, i.e. it is not
self-consistent. This can cause similar methods to lead to different
results due to slightly different LDA Green functions as was also
recognized when the LDA Green function was renormalized with the GW
term \cite{Schilfgaarde2006}. The obvious way to overcome this problem
is to impose some kind of self consistency, which is a work in progress. In our method, a self-consistent scheme can be constructed by using the calculated self-energy to calculate new interaction parameters $J$ which are then used to calculate the renormalized Heisenberg susceptibility, leading to a new self-energy. The implementation of this scheme is still a work in progress.  

\section{Results and Discussion}\label{chap_res}

\subsection{bcc iron}
\begin{figure}
    \centering
    \includegraphics[width=0.45\textwidth]{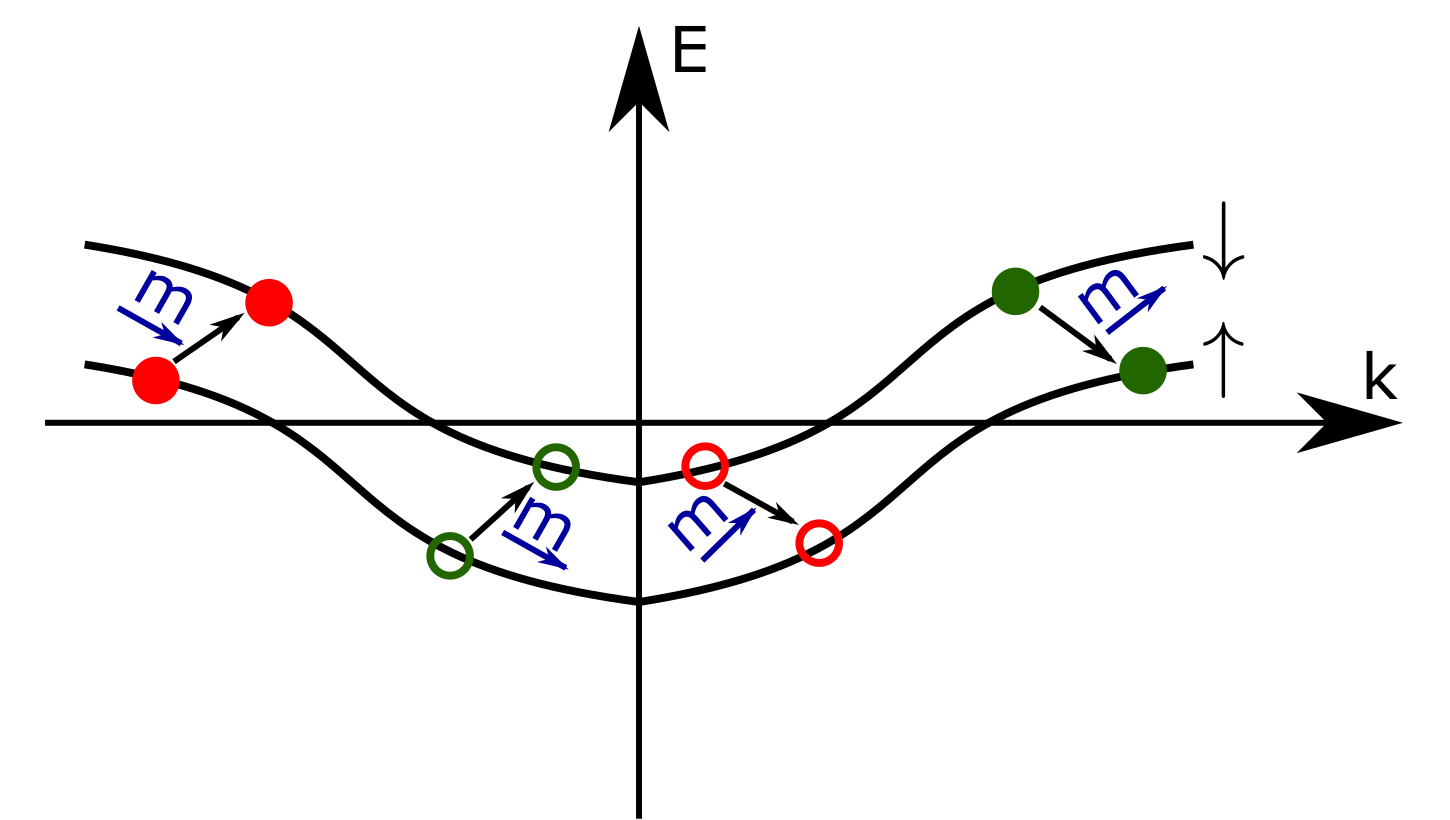}
    \caption{Different quasiparticle transitions involving magnons in a simplified band structure. A collinear magnet without spin-orbit interaction at low temperatures is considered. As there are no thermally excited magnons, the magnon emission processes dominate. In the strong ferromagnets, the magnons are mainly associated with the decrease of the angular momentum (they are, effectively, up-to-down spin-flips). From left to right: A majority electron absorbs a magnon (blue arrow labeled m) and gets excited into an empty minority spin state. A majority hole is excited into a minority hole by emitting a magnon. A minority hole decays into a majority hole by absorbing a magnon. A minority electron decays into a majority state and emits a magnon in the process. At low temperatures, the processes colored green are dominant while the processes colored red are negligible. There are further thinkable spin-dependent processes: the emission and absorption of spin excitations which increase the total spin angular momentum (involving, effectively, down-to-up spin-flips). At low temperatures, they could be emitted by an up electron or down hole. In the ferromagnets discussed in this paper, these so called anti-Stoner processes feature negligible spectral weight and are not considered.
}
    \label{fig_processes}
\end{figure}

The spectrum of iron under the influence of the electron-magnon
interaction is presented in figure \ref{fig_Fe} a and b. Generally, the
electron-magnon interaction leads to a broadening of the electron
states as well as a renormalization of the electron energies and the
appearance of additional peaks. Our result agrees qualitatively with
recently published results for iron using an the MBPT method
\cite{Mueller2019,Nabok2021} and the DMFT methods
\cite{Katsnelson1999,Braun2006,Grechnev2007,SanchezBarriga2009,SanchezBarriga2012}. All
of the aforementioned works report larger broadening for majority
spin carriers, especially around $E=-2$ eV across the whole
spectrum. In the minority spin channel we observe an additional peak
at $\Gamma$ for $E\approx0.5$ eV, which does not appear in any of the
other works. The spectral weight of this peak is however very small
and hence, it might be due to a slightly different LDA Green function
compared to \cite{Mueller2019}. It is certainly not caused by the
magnon-pole approximation as will be shown in the following. Regarding
the renormalization of the electronic energies compared to the LDA
solution, our results show the same trend that was found in
\cite{Mueller2019} and in the $GT$ results of \cite{Nabok2021}. The different absolute values are partly due to
the magnon-pole approximation and can also arise due to a slightly
different LDA Green functions. \newline
While the majority spin electrons are shifted to higher energies, the minority bands remain mostly at the same energies. Due to the fact that several majority bands cross the LDA Fermi energy, the renormaliztation is expected to also have an impact on the Fermi energy. The change of the Fermi energy can be calculated from the renormalized density of states (DOS) by ``filling up'' the states until the correct number of electrons is reached. It needs to be mentioned at this point that the calculation of the renormalized DOS is only feasible when the self-energy is assumed to be constant in $\vb*{k}$, which is not quite the case for iron as will be discussed in more detail later. Hence, the resulting changes of the Fermi energy (and the magnetic moment) for the case of iron should be viewed as an approximation. The green dashed line in figure \ref{fig_Fe} a and b indicates the new Fermi level which lies about 0.27eV above the LDA Fermi energy. The magnetic moment per atom is reduced to $\mu=2.05\mu_{\text{B}}$ giving a slight deviation from the experimental result of $\mu\approx2.2\mu_{\text{B}}$ \cite{Bardos1969} which is accurately predicted within the LDA. The renormalization of the Fermi energy and the magnetic moments is significantly higher for iron than for any other material studied here. It may be speculated that a self-consistent procedure is more important in this case.   \newline
Note that the spectrum of spin up electrons is mostly affected below
the Fermi energy while for spin down electrons the opposite trend can be
observed. This is caused by the imaginary part of the self-energy being approximately zero above (below) the Fermi energy for minority (majority) spin carriers. 
Physically, this asymmetry arises from the extreme case of a strong ferromagnet at zero temperature. In figure \ref{fig_processes} all possible quasiparticle transitions in a strong ferromagnet are schematically drawn. Above the Fermi energy a minority spin electron can decay into a majority spin state by emitting a magnon. The opposite process of a majority spin electron being excited into a minority state does not occur as it would need a (e.g. thermally excited) magnon to account for spin conservation. The corresponding argumentation applies for holes below the Fermi energy.\newline
The damping of spin-waves through hybridization with the Stoner continuum is an important mechanism in materials with itinerant electrons. Especially in iron this effect was shown to lead to ``spin-wave disappearance'' \cite{Buczek2011}. Thus, although most of the spectral weight of the magnetic spectrum lies in the spin-wave region, the assumption of sharp magnon peaks is generally not fulfilled throughout the Brillouin zone. Hence, we study the influence of the magnon-pole approximation for a small energy range at the $\Gamma$ point and minority carriers, as shown in figure \ref{fig_Fe} c. 
The LDA result is compared with the renormalized Green functions
computed within the magnon-pole approximation (LDA+$\mathcal{V}_{e-m}$) and with the TDDFT
susceptibility (LDA+$\mathcal{V}_{e-m}^{\text{TDDFT}}$). The previously discussed peak is visible for both
Green functions renormalized with the TDDFT susceptibility and its
Heisenberg counterpart. The position of the peaks differs slightly and
the peaks calculated with the TDDFT susceptibility are slightly
broader. However, we conclude that the major effects of the
renormalization are properly included within the magnon-pole
approximation.\newline
Over recent years several experimental studies of the band structure of
iron were conducted
\cite{SanchezBarriga2009,SanchezBarriga2012,Schaefer2005}. In all
these studies the effective mass renormalization
$\frac{m^\star_\alpha}{m^\star_{0\alpha}}$ was measured by means of
angle resolved photoemission spectroscopy
(ARPES).
Using DFT and ARPES quasiparticle energies, Sanchez-Barriga
et. al. \cite{SanchezBarriga2009,SanchezBarriga2012} derive an energy
dependence of the self-energy from which they calculate the effective
mass renormalization and compare to DMFT results. They find that in
bcc iron the effective mass renormalization depends strongly on the
position in the Brillouin zone.
However, the DMFT cannot account for a $\vb*{k}$ dependence in the self
energy and hence gives the same result independent of the position in
the BZ. While they show that this is a good approximation for Ni the
situation is different in the case of Fe. Our results for the trace of the self-energy
calculated for iron (figure \ref{fig_Fe} d) and nickel (figure \ref{fig_Ni} e) at different positions in the BZ show that for nickel the self-energy is practically constant while for iron a clear dependence on the position in the BZ can be observed. This is also manifested in the values for the effective mass renormalization which we find to vary between 1.54 at the N point and 2.25 at the $\Gamma$ point for the $\Sigma_4$ band.
Hence, our results confirm that the non-local character of electron
correlations in bcc iron is essential for the proper band structure
description. The physical origin of this difference between Fe and Ni 
will be discussed below, at the end of the next subsection. \newline
Schaefer et. al. \cite{Schaefer2005} compare the slopes
of bands at the Fermi energy instead (i.e. for constant energy instead of
constant $\vb*{k}$). As reference they use their own DFT calculations
which, although qualitatively similar, differ from our results. To
eliminate the dependence on the DFT reference we compare the velocity
of the renormalized quasiparticle obtained in \cite{Schaefer2005}
directly from the experimental data, cf. table \ref{tab_mass2}. We
calculate the velocity by taking the derivative of the quasiparticle
energy (i.e. the position of the peak maximum) with respect to
$\vb*{k}$. Apart from point I, the electron-magnon interaction reduces the quasiparticle velocity compared to the LDA which leads to a better agreement with the experimental data.
\begin{table}
	\begin{tabular}{cc|c|c|c|}
		& & Experiment \cite{Schaefer2005} (eV\AA)&\multicolumn{2}{c}{Our result (eV\AA)}\\
		&&&LDA+$\mathcal{V}_{e-m}$&LDA\\
		\hline
		\multirow{3}{*}{Spin $\uparrow$}
		&$\Gamma$-H (I)&1.12&2.2&2.2\\
		&$\Gamma$-N (II)&1.16&1.2&3.5\\
		&$\Gamma$-P (III)&1.4&2.3&4.3\\
	\end{tabular}
	\caption{Velocity of quasiparticles
          $v=\pdv{\varepsilon}{\vb*{k}}$ at the Fermi energy in bcc
          iron from the experimental study \cite{Schaefer2005}
          compared with our results. The labels next to the direction
          refer to the points in the spectrum (figure
          \ref{fig_Fe} a and b).} 
	\label{tab_mass2}
\end{table}
\subsection{fcc Ni}
\begin{figure*}
    \includegraphics[width=\textwidth]{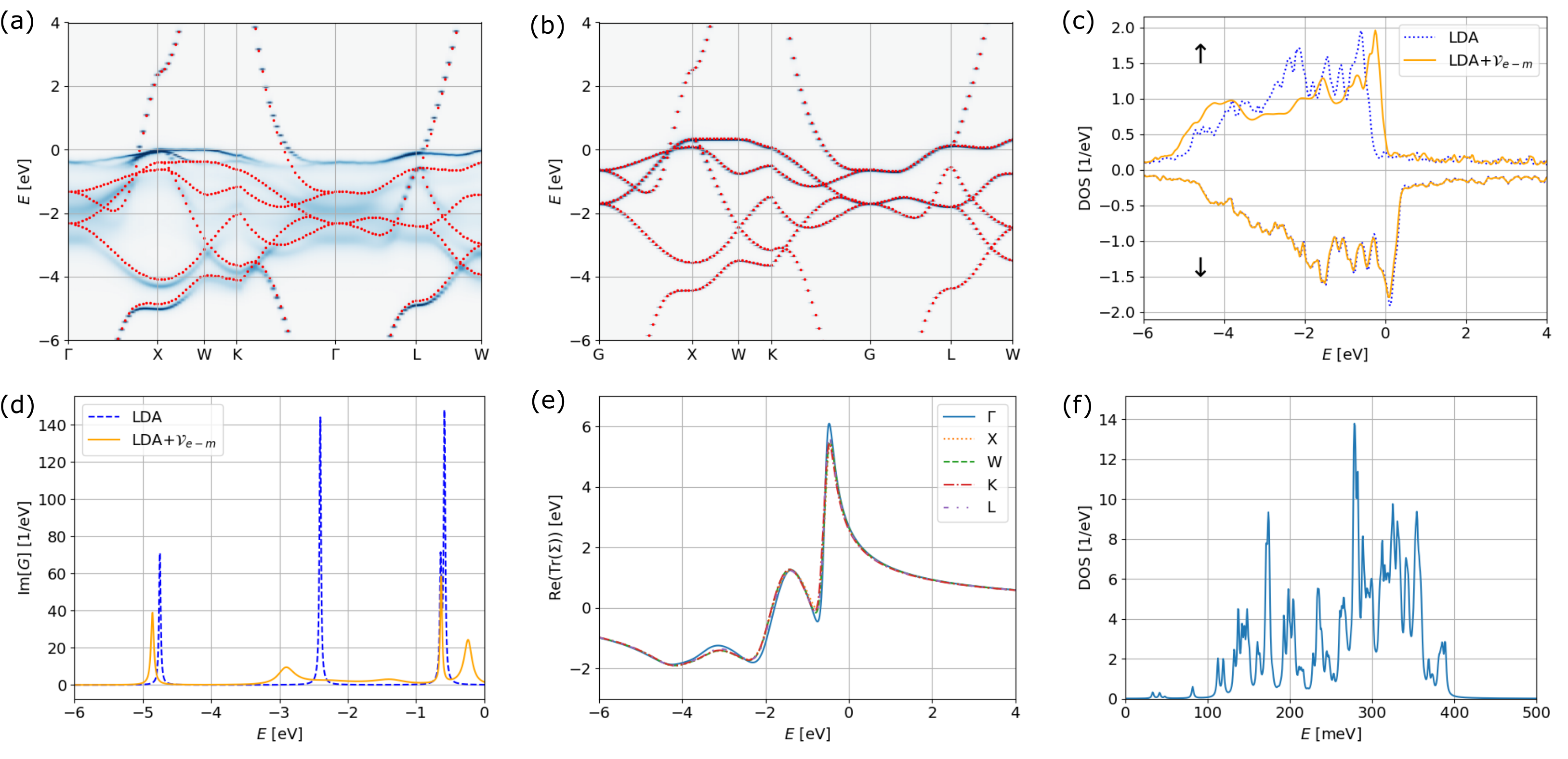}
    \caption{Effect of the electron-magnon interaction in fcc nickel. (a) Electronic band structure of majority carriers under the influence of the electron-magnon interaction. The red dots represent the energy levels within the LDA while the blue background indicates the renormalized spectrum. (b) Same as (a) for minority carriers. (c) Electronic density of states within the LDA and the LDA under the influence of the electron magnon scattering (LDA+$\mathcal{V}_{e-m}$). (d) Spectral function at the L point from the LDA and the renormalized Green function. (e) Real part of the trace of the self-energy for different points in the Brillouin zone. (f) Magnonic density of states.}
    \label{fig_Ni}
\end{figure*}
\begin{table}
    \centering
    \begin{tabular}{c|c|c|c|c|c}
         &LDA&LDA+$\mathcal{V}_{e-m}$& $GWT$& LDA+DMFT&exp. \\\hline
         L$_3$&0.71&0.31&0.37&$\sim $0.4&0.31$\pm$0.03\\
         X$_2$&0.7&0.23&0.31& &$ \sim $ 0.2
    \end{tabular}
    \caption{Comparison of exchange splittings in nickel (in eV) with the GWT approach \cite{Nabok2021}, the LDA+DMFT \cite{SanchezBarriga2012} and experimental results \cite{Himpsel1979,Raue1984}.}
    \label{tab_Exc_comparison}
\end{table}
Nickel is an example of a material not well described by the
LDA \cite{Lichtenstein2001}. Here, the most prominent features are the missing satellite peak
at 6 eV binding energy, a far too large exchange splitting and a too
broad valence band. These discrepancies are believed to be a result of
many-body effects, which cannot be described within the LDA, in
particular those arising from the coupling to spin fluctuations. The DFT+DMFT solves the problem and allows to reach a very good agreement 
with the experiment eliminating all three discrepancies \cite{Lichtenstein2001,Braun2006,SanchezBarriga2012}.
At the same time, the DMFT takes into account different kinds the many-body
processes. It is instructive to separate the effects of the electron-magnon 
interaction, which we will do here, without local approximation for the electron
self-energy. \\
We find that the 6 eV satellite of Ni is not originated by coupling to spin-flip excitations.  Also, the narrowing of the valence band width is not a
consequence of the electron-magnon scattering. In fact, the bandwidth
is slightly increased for the renormalized Green function, cf. figure
\ref{fig_Ni} a and d. Thus, from these three problems of the LDA, only one, namely,
the overestimate of the spin splitting, is related to electron-magnon
processes and is fully grasped in our approach. 

Generally, the majority spin bands are strongly
damped by the electron-magnon interaction as can be seen by the blue
background in figure \ref{fig_Ni} compared to the LDA solution
indicated by red filled dots. The minority spin carriers in contrast
are hardly influenced by the electron-magnon interaction, cf. figure \ref{fig_Ni} b. This is due
to the spin asymmetry of the density of states, as also reported in
\cite{Mueller2019}.
We note that the exchange splitting is strongly reduced by a new peak in
the majority spin channel appearing throughout the whole spectrum
close to the Fermi energy visible in figures \ref{fig_Ni} a and
\ref{fig_Ni} d. Our results for the exchange splitting at the $L$ and $X$ points are in excellent agreement with ARPES results, as shown in table \ref{tab_Exc_comparison}. 
The shift of majority electron states towards the Fermi
energy as a consequence of the electron-magnon interaction is reflected clearly in the DOS, as shown in figure \ref{fig_Ni} c. 
However, in the ARPES experiments no additional peak (compared to LDA
results) was found. The reason for this can be seen in figure
\ref{fig_Ni} d which shows the spectral weight (which is proportional
to the imaginary part of the Green function) at the $L$ point for the
LDA and renormalized Green functions. The LDA results include a peak
at -2.5eV which is damped by the strong electron-magnon interaction to such
an extent that it can hardly be identified as a well defined
quasiparticle any more. Consequently, our results only feature three
clear peaks in agreement with the ARPES studies.  \newline
Similar to iron, a shift of the majority spin bands towards higher energies can be observed in figure \ref{fig_Ni} a. However, here the renormalized bands are still mostly located below the Fermi energy. Consequently, the renormalization has practically no effect on the Fermi energy (it shifts by less than 30meV) and the magnetic moment (it decreases by approximately 0.072$\mu_{\text{B}}$).\newline

As was already mentioned at the end of the previous section, the self-energy
in nickel is much less dependent on the wavevector than in iron, in agreement
with the previous conclusions based on the comparison of DFT+DMFT data with 
the experiment \cite{Lichtenstein2001,SanchezBarriga2012}. 
The difference might
be associated to an essential difference of exchange interactions and thus magnon
spectra between bcc Fe and fcc Ni. Already from the values of the spin-wave
stiffness constant it is clear that magnons in Fe are much softer than in
Ni despite the fact that the magnetic moments of the atoms are almost four time larger in the
former case than in the latter. This is related with two circumstances.
First, Friedel oscillations of exchange parameters are much more pronounced
for Fe than for Ni \cite{Pajda2001,Kvashnin2015}. The reason for this is that Ni is
almost half-metallic, with almost fully occupied majority-spin band whereas
the Friedel oscillations require metallicity for both spin projections. Second, the magnetic interactions in Fe are frustrated, with the ferromagnetic
e$_g$ electron subsystem and the antiferromagnetic t$_{2g}$-t$_{2g}$ contributions
to the exchange parameters \cite{Kvashnin2016}. On the opposite, magnetism
of Ni is mostly determined by the nearest-neighbor ferromagnetic interactions
without any essential competing interaction. 

Let us conclude this section with a short comment on the electron-magnon scattering in the spin-polarized uniform electron gas at densities corresponding to the $3d$ transition metals as it can provide us with an additional insight into the emergence of the magnetic order in these systems. A simple model of these magnets involves two parabolic bands split by a uniform exchange-correlation magnetic field $B_{\text{xc}}$. The Fermi energy $E_{\text{F}}$ and $B_{\text{xc}}$ are chosen such that the charge density equals to the one of the valence band of a $3d$ system and the magnetic moment assumes the corresponding value. This model yields a qualitatively correct picture of the magnons in itinerant systems \cite{Moriya2012} with a magnon band featuring the Goldstone mode and a quadratic dispersion for small wave vectors, and interacting with the Stoner continuum. Following our scheme, also the self-energy arising due to the exchange of virtual magnons can be computed. It turns out, however, that its value is much larger than $E_{\text{F}}$ and would destroy the magnetic order thus yielding a qualitatively incorrect picture of magnetism in this model. The failure is attributed to the unrealistically large exchange splitting $B_{\text{xc}}$, and correspondingly the exchange-correlation kernel $K_{\text{xc}}$, necessary to induce the magnetic moment according to the Stoner criterion ($K_{\text{xc}}\chi_0=1$). In real $3d$ transition metals, the emergence of the magnetic order arises due to the large value of the non-interacting susceptibility $\chi_0$ being in turn proportional to the large density of states in the partially localized 3d band. Thus, the real $K_{\text{xc}}$ coupling is small and the self-energy, as expected, turns out to be a correction to the band structure.

\subsection{NiMnSb}
\begin{figure*}
    \includegraphics[width=\textwidth]{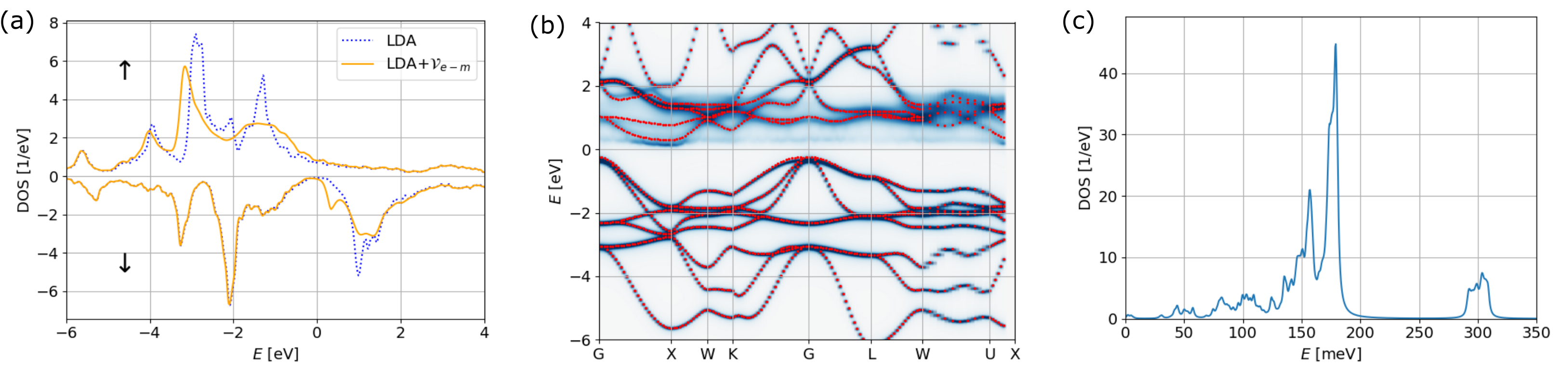}
    \caption{Effect of the electron-magnon interaction in NiMnSb. (a) Electronic density of states within the LDA and the LDA under the impact of the electron magnon scattering (LDA+$\mathcal{V}_{e-m}$). (b) Electronic band structure of minority carriers under the influence of the electron-magnon interaction. The red dots represent the energy levels within the LDA while the blue background indicates the renormalized spectrum. (c) Magnonic density of states.}
    \label{fig_NiMnSb}
\end{figure*}
NiMnSb is seen as a prototype example for a half-metallic ferromagnet (HMF), i.e. it is
conducting for one spin direction and a semiconductor for the other \cite{Groot1983,Katsnelson2008}.
In the case of NiMnSb the LDA density of states near the Fermi energy is
zero for the minority spin channel as depicted in figure
\ref{fig_NiMnSb} a. Prediction of the half-metallicity in this
compound \cite{Groot1983} opened the field of half-metallic ferromagnetism.

The use of the magnon-pole approximation is particularly justified in
half-metallic ferromagnets. The gap in the electronic spectrum at the
Fermi level causes the Stoner excitations to emerge at energies much
larger than typical for spin-waves. The latter are therefore
Stoner undamped and contain the majority of the spectral weight \cite{Katsnelson2008,Buczek2009}.

The emergence of non-quasiparticle states (NQPS) in the spin gap is a
trademark of half-metallic ferromagnets. The physical picture of such
states is that spin down states directly above the Fermi energy can be
seen as a superposition of spin up states and virtual magnons
\cite{Irkhin1990,Chioncel2003,Katsnelson2008}. Contrary to quasiparticle states, which arise from
the poles of the Green function, NQPS are caused by its branch
cuts \cite{Irkhin1990}. There is no spectral weight of such states
directly at the Fermi energy but it increases drastically at energy
scales of characteristic magnon frequencies. The spectrum of minority
spin carriers, cf. figure \ref{fig_NiMnSb} b, shows a strong
broadening of electronic states above the Fermi energy. Across the
whole spectrum a dispersionless peak appears approximately
150~meV-300~meV above the Fermi energy. The magnonic density of states
\cite{Paischer2021,Paischer2021a}, cf. figure
\ref{fig_NiMnSb} c confirms a high number of magnon states
within that energy region (the electronic DOS of majority carriers is
approximately constant in the relevant energy window, cf figure \ref{fig_NiMnSb} a). 
Note that this density of states looks much simpler than that for Fe (Fig. \ref{fig_Fe} e) and Ni
(Fig. \ref{fig_Ni} f), with just a few well-pronounced main peaks.  
Similar to the
case of nickel, the self-energy of NiMnSb depends only weakly on
the position in the BZ which makes our results quite close to results obtained from the DFT+DMFT
\cite{Chioncel2003}. This is a general feature of the self-energy
for most materials as the magnon energies are small compared to the
typical electronic bandwidth. As we already discussed, bcc Fe seems to be a
counterexample. However, in this case its ferromagnetism is related to
e$_g$ states only, and one can assume that instead of the total bandwidth
it is the width of this peak which matters \cite{Irkhin1993,Kvashnin2016}.

Hence, the electronic density of states
may be calculated under the assumption that the self-energy is
independent of $\vb*{k}$ hence lowering the numerical burden of the
calculation. Our results, presented in figure \ref{fig_NiMnSb} a, show a
broadening of the sharp peaks in the DOS caused by the broadening of
the electron states through the electron-magnon interaction. 
In addition, a peak in the DOS emerges
in the gap in the minority spin channel due to the NQPS formation. \newline
Similar to the situation in nickel, the Fermi energy as well as the magnetic moments of the system remain practically unchanged through the renormalization (the Fermi energy increases by 86meV and the total magnetic moment increases by 0.001$\mu_{\text{B}}$).



\section{Summary and outlook}
In this work we presented an ab initio method to account for the influence of non-local correlations due to spin-flip processes on the electronic structure of ferromagnets. It is based on the MBPT in the formalism of Hedin \cite{Hedin1965} but avoids its main disadvantage, i.e. the high numerical complexity, as the main quantities used are approximated using TDDFT. This methods gives very similar results to other more complex methods \cite{Mueller2019,Nabok2021} for bcc iron and fcc nickel. Most notably, the predicted exchange splitting in fcc nickel is closer to the experimental value than those of other comparable methods. The lower numerical cost allows to study complex materials as the half-metallic ferromagnet NiMnSb. In the latter material, the electron-magnon interaction causes the appearance of non-quasiparticle states in the bandgap. Contrary to DMFT, we are able to account for the non-locality of the self-energy. Extensions of the theory towards antiferromagnets and 2D systems as well as the influence of paramagnons are a current work in progress.




\section*{Acknowledgments}
S.P. is recipient of a DOC Fellowship of the Austrian Academy of Sciences at the Institute of mathematics, physics, space research and materials sciences. P.A.B. and A.E. acknowledge the funding by the
Fonds zur F\"orderung der Wissenschaftlichen Forschung (FWF) under Grant No. I 5384. P.A.B. acknowledges the generous support of the Alexander von Humboldt Foundation during his stay as a Feodor Lynen Research Fellow at the University of Missouri–Columbia, USA. The work of M.I.K. was supported by the European Union’s
Horizon 2020 research and innovation programme under
European Research Council synergy grant 854843 “FASTCORR”. sponding authors upon request.

\bibliographystyle{apsrev4-2}
\bibliography{./Quellen}

\newpage\newpage

\end{document}